\documentclass[review]{elsarticle}
\usepackage{lineno,hyperref}
\usepackage{epsfig}
\usepackage{natbib}
\usepackage{amsfonts}
\usepackage{amsmath}
\usepackage{graphicx}
\usepackage{amssymb}
\usepackage{epsfig}
\usepackage{tabularx}
\usepackage{longtable}
\usepackage{adjustbox}
\usepackage{supertabular}
\usepackage{booktabs}
\usepackage{caption}
\usepackage{ctable}
\usepackage{float}
\usepackage{url}
\usepackage{lscape}
\usepackage{multirow}
\usepackage{subcaption}
\usepackage[font=scriptsize]{caption}

\modulolinenumbers[5]
\makeatletter
\def\ps@pprintTitle{%
  \let\@oddhead\@empty
  \let\@evenhead\@empty
  \let\@oddfoot\@empty
  \let\@evenfoot\@oddfoot
}
\makeatother

\begin{document}

\begin{frontmatter}
\title{Forecasting time series with structural breaks with\\
Singular Spectrum Analysis, using a general form of\\
recurrent formula}

\author[mymainaddress]{Donya Rahmani}
\address[mymainaddress]{Faculty of Science and Technology, Bournemouth University }
\ead{drahmani@bournemouth.ac.uk.}

\author[myfirstcoauthermainaddress]{Saeed Heravi}
\address[myfirstcoauthermainaddress]{The Business School, Cardiff University}

\author[mysecondcoauthermainaddress]{Hossein Hassani}
\author[mysecondcoauthermainaddress]{Mansi Ghodsi}
\address[mysecondcoauthermainaddress]{The Business School, Bournemouth University}


\begin{abstract}
This study extends and evaluates the forecasting performance of the Singular Spectrum Analysis (SSA) technique using a general non-linear form for the recurrent formula. In this study, we consider 24 series measuring the monthly seasonally adjusted industrial production of important sectors of the German, French and UK economies.  This is tested by comparing the performance of the new proposed model with basic SSA and the SSA bootstrap forecasting, especially when there is evidence of structural breaks in both in-sample and out-of-sample periods. According to root mean-square error (RMSE), SSA using the general recursive formula outperforms both the SSA and the bootstrap forecasting at horizons of up to a year. We found no significant difference in predicting the direction of change between these methods. Therefore, it is suggested that the SSA model with the general recurrent formula should be chosen by users in the case of structural breaks in the series.
\end{abstract}

\begin{keyword}
State Dependent Models \sep Singular Spectrum Analysis \sep Forecasting \sep Industrial Production.
\MSC [2010] 37M10\sep 91B84
\end{keyword}

\end{frontmatter}



\section{Introduction}
Singular Spectrum Analysis (SSA) is a powerful method of non-parametric time series analysis and forecasting. In the last 20 years, there has been much interest to develop and apply this technique to a wide range of applications, ranging from mathematics and signal processing to meteorology, economics and market research (see, for example \cite{Ghil2002}, \cite{Moskvina2003} and \cite{Hassani2007}). The SSA method is based on decomposing the time series into the three components of trend, harmonics and noise. The method then reconstructs the original series and computes the forecasts based on the reconstructed series. The main advantage of SSA is the ability to forecast a series which is not normally distributed and have complex seasonal components and non-stationary trend. Thus the method can be used without making statistical assumptions such as stationarity and normality of the data and residuals. A thorough description of the theoretical and practical foundations of the SSA method can be found in the books by \cite{Elsner1996} and \cite{Nina2001}.

Index of Industrial Production (IIP) and its components are the most widely used time series by policymakers and economists and therefore their accurate forecasts are of great importance. Franses and Van Dijk \cite{Franses2005} examined the forecasting performance of various models, using quarterly industrial production data of 18 countries; they found that ARIMA models generally performed well in the short-term out-of-sample forecast. However, for longer horizons non-linear, more complex models provide more accurate forecasts. They concluded that none of the methods employed in their study gave totally satisfactory results and thus they advised the use of forecast combination. Heravi et al. \cite{Heravi2004} compared the performance of Artificial Neural Networks and linear forecasts, using seasonally unadjusted monthly industrial production indices for eight important sectors of the economies of the United Kingdom, France and Germany. They found some evidence of non-linearity in the data and concluded that non-linear models, such as neural networks, dominate linear models in the prediction of direction of change but not in actual forecasting performance at horizons of up to a year. Hassani et al. \cite{Hassani2009} extended the data on industrial production used by \cite{Heravi2004} and compared the performance of SSA with that of Holt-Winters and ARIMA. Based on RMSE, they concluded that the three methods performed similarly for short term forecasting, but SSA performed better than ARIMA and Holt-Winters’ models at the longer horizons. They also noted that SSA performed well for short time series and that all three methods similarly predicted  well the direction of change points. Dahl and Hylleberg \cite{Dahl2004} evaluated the out-of-sample accuracy of non-linear models and compared them with linear models using various measures, concluding that in general, non-linear models outperformed the linear models. Patterson et al. \cite{Patterson2011} also applied and showed the benefits of using the Multivariate SSA for real time forecasting revisions on the UK index of industrial production data. A full review of the application of SSA for Economic and Financial time series is given in \cite{Hassani2010}.

Priestley \cite{Priestley1980} developed a general class of time series model called state-dependent models. This model includes non-linear as well as the standard linear time series models. In this study we mainly follow the method used in \cite{Priestley1980} and extend the Singular Spectrum Analysis technique by considering a more general form of the recurrent formula. We update the parameters obtained by the optimal SSA (bootstrap SSA) in the out-of-sample forecast period, relating them to past values of the observed process. This study examines the impact of updating the parameter estimates within the forecast period.

The paper is organized as follows: In section 2, a brief description of the Singular Spectrum Analysis technique and the bootstrap SSA are given. Section 3 extends the SSA and describes the algorithm for updating the coefficients based on the general recurrent formula. Section 4 outlines the data used in the study and apply Bai and Perron test to detect  the structural breaks in the data. Section 5 presents the empirical results and discusses the findings. Some conclusions are drawn in section 6.

\section{SSA and Bootstrap SSA}
This section gives a brief description of  Singular Spectrum Analysis. SSA is a method that decomposes the original series into a number independent components, namely, trend, periodic and noise components. SSA consists of two stages, decomposition and reconstruction. The first stage decomposes the times series and the second stage reconstructs the decomposed series and obtain forecasts via a linear recurrent formula.
\subsection{Decomposition}
In the decomposition stage, SSA first organizes the one dimensional times series data into a multidimensional series, by selecting a vector of $L$ observations and moving this throughout the sample. This procedure is called embedding and results in the trajectory matrix, $\mathbf{X}$, with dimensions of L by $K = N-L + 1$
{\setlength\arraycolsep{.3em}
\begin{eqnarray*} \mathbf{X} ={ \begin{pmatrix} y_{1}&y_{2}&\cdots&y_{N-L}&y_{N-L+1}\\ y_{2}&y_{3}&\cdots&y_{N-L+1}&y_{N-L+2}\\ \vdots  & \vdots  & \vdots & \vdots & \vdots  \\  y_{L-1}&y_{L}&\cdots&y_{N-2}&y_{N-1}\\ y_{L}&y_{L+1}&\cdots&y_{N-1}&y_{N}\\ \end{pmatrix} }.\\ \label{eq:trajectory} \end{eqnarray*}}
The ``window length" $L$ is an integer between 2 and $N$ and needs to be set in the decomposition stage. Choice of  $L$ depends on the structure of the data and no single rule for $L$ can cover all applications. However, in general $L$ should be proportional to the periodicity of the data, large enough to obtain sufficiently separated components but not greater than $N/2$. Full discussion of the choice of this parameter is given in \cite{Nina2001}.

 In the second step, the trajectory matrix $\mathbf{X}$ is decomposed into the sum of $d$ elementary matrices:
\begin{equation}
\mathbf{X}=\mathbf{X}_1+\ldots+\mathbf{X}_d,
\label{SVD1}
\end{equation}
where, $\mathbf{X}_i=\sqrt{\lambda_i}U_iV_i^T$ (for $i=1,\ldots,d$) and  $\lambda_1,\lambda_2\ldots,\lambda_d$ are ordered nonnegative eigenvalues of $\mathbf{S}=\mathbf{XX}^{T}$, and $U_1,\ldots,U_d$ are the corresponding eigenvectors of $\mathbf{S}$. The principal components  are computed by $V_i=\mathbf{X}^TU_i/\sqrt{\lambda_i}$ and the set $(\sqrt{\lambda_i},U_i,V_i)$ are called the $i^{\text{th}}$ eigentriple of the matrix $\mathbf{S}$.

\subsection{Reconstruction}
To distinguish the signal from the noise, we need to separate the elementary matrices into two sets of  $I=\{1,2,\ldots,r\}$ and its complement  $\bar{I}=\{r+1, r+2,\ldots,d\}$. The first $ r$ elementary matrices  $\mathbf{X}_1 ,\ldots , \mathbf{X}_r$  which approximate the original matrix $\mathbf{X}$ are used to construct the signal, the rest considered as noise. The contribution of the first $r$ elementary matrices $\mathbf{X}_1 ,\ldots , \mathbf{X}_r$  is measured by the share of the corresponding eigenvalues, ${\sum_{i=1}^{r}\lambda_i}/{\sum_{i=r+1}^{d}\lambda_i}$.

A reconstructed series of the same length as the original series can be obtained by performing the diagonal averaging (Hankelisation) of the matrix   $\mathbf{X}_1+\ldots+\mathbf{X}_r$. In the reconstruction stage the number of elementary matrices, $r$, should be selected. This parameter can be selected by looking for the breaks among the eigenvalues $\lambda_1,\lambda_2\ldots\lambda_d$ and by the periodogram analysis of the original time series. For detailed discussion of the selection of this parameter see \cite{Nina2001,Hassani2009}.

\subsection{Forecasting}
In order to perform forecasting with SSA, the time series has to approximately satisfy the linear recurrent formula given by
\begin{equation}
y_t=\sum_{j=1}^{L-1}\phi_jy_{t-j},
\label{coeff}
\end{equation}
where vector $\{\phi\}_{j=1}^{L-1}$ is a sequence of constant coefficients and should be determined. In SSA the recurrent coefficients  $\phi_{j}, j=1,\ldots,{L-1}$, are obtained as follows.
First define $A = (\phi_{L-1},\phi_{L-2},\ldots,\phi_1)$; next distinguish the first $L-1$ components of the eigenvector $U_i$, such that   $U_i=({U_i}^{\nabla L},\pi_i)$ where  ${U_i}^{\nabla L}=(u_{i,1},\ldots,u_{i,L-1})$ and $\pi_i$ is the last component of $U_i$; and define $\Pi=\sum_{i=1}^{r}\pi_i^2<1$. The recurrent coefficients are then obtained as:

\begin{equation*}
A={(1-\Pi)^{-1}} \sum_{i=1}^{r} \pi_i {U_i}^{\nabla L}\label{R}.
\end{equation*}

Once the recurrent coefficients $\phi_j$ , $j = 1,\ldots, L- 1$ are determined, the dynamic forecasts can be simply obtained by applying the recurrence formula.
 \begin{equation}
\hat{y}_{t+h}=\sum_{j=1}^{L-1}\phi_j\hat{y}_{t+h-j}\qquad h=1,2,3,\ldots
\label{coeff2}
\end{equation}

Within the estimation sample $\hat{y}_{t}$, the fitted values of  ${y}_{t}$, are just the same as the reconstructed series $\tilde{y}_{t}$; $\hat{y}_t=\tilde{y}_{t}$ for $t=1,2,\ldots,N $.

\subsection{Bootstrap SSA}\label{sec:boot}
A forecasting method can be assessed either by Monte Carlo simulations or bootstrap. However, the Monte Carlo simulations can be applied in situations when the true model is known. In SSA, the true model for the signal is not known before filtering and reconstruction, and thus the bootstrap procedure is applied to obtain the statistical properties of the forecasts and construct interval estimates.

Assume that we have a time series which consists of two components $Y_N=\{y_t\}_{t=1}^{N}= S_N+E_N$ where $S_N$ is the signal and $E_N$ is the noise.
Under a suitable choice of the window length $L$ and the corresponding eigentriples, there is a representation $\hat{Y}= \hat{S}_N+\hat{E}_N$, where $\hat{S}_{N}$ (the reconstructed series) approximates $S_N$ and $\hat{E}_{N}$ stands for the residual series. Through bootstrap technique  (with replacement), we generate $B$ independent ``copies"  $E_{N,i} (i=1,\ldots,B)$  of $\hat{E}_N$ from the noise.  We can then obtain $B$ series $\hat{Y}= \hat{S}_N+\hat{E}_N$ and produce $h$ forecasting $\hat{S}_{N+h}$ in the same manner as in the Monte Carlo simulation.

Average bootstrap forecasts can then be computed from the sample  $\hat{S}_{N+h,i}\\(1 \leq i \leq N )$ of these forecasts $\hat{S}_{N+h}$ and be compared with the forecasting results obtained by the basic SSA. Large discrepancy between these two forecasts would typically indicate that the original SSA forecasts are not reliable.

\section{SSA Using a  General Recurrent Formula }
Priestley  \cite{Priestley1980} developed a general class of time series models, called “State Dependent Models” (SDM), which includes non-linear time series models and standard linear models as special cases. The principal advantage of SDM is that it is flexible and can be fitted without any specific prior assumption about the parameters. This is not only useful in itself, but may give an indication of specific type of non-linearity and structural breaks in the data, or even, indeed, whether a linear model with constant coefficients might prove to be equally satisfactory. A more extensive discussion and identification of these general models is given in \cite{Priestley1980}. An extensive study of the application of state-dependent models to real and simulated data is given in \cite{Haggan1984} and its extension to non-linear dynamical systems is studied by \cite{Priestley1986}.
In this section we extend SSA with the linear recurrent formula to a general state dependent form. We explain how the estimated coefficients of the linear recurrent formula obtained by SSA within the sample period can be recursively updated in the forecast period.
Consider the following  linear recurrent formula given in equation (\ref{coeff})
\begin{equation}
y_t=\phi_1 y_{t-1}+\phi_2 y_{t-2}+\ldots+\phi_{L-1} y_{t-(L-1)},
\label{AR}
\end{equation}
where \{$\phi_1,\phi_2,\ldots,\phi_{L-1}\}$  are constant and obtained by SSA within the sample period,  then at time $(t-1)$ the future development of the process $\{y_t\}$ is determined by the values $\{y_{t-1},\ldots,y_{t-(L-1)}\}$.  Hence, the vector $Y_{t-1}=\{y_{t-1},\ldots, y_{t-(L-1)}\}^T$  may be regarded as the ‘state-vector’ of the process $\{y_t\}$. That is, the only information in the ‘past’ of the process relevant to the future development of the process which is contained in the state-vector. The SDM extends the idea of the linear recurrent formula by allowing the coefficient of model (\ref{AR}) to become functions of the state-vector $Y_{t-1}$, leading to the general recurrent formula model:
\begin{equation}
y_t=\phi_1(Y_{t-1}) y_{t-1}+\phi_2(Y_{t-1}) y_{t-2}+\ldots+\phi_{L-1}(Y_{t-1}) y_{t-(L-1)},
\label{SDMAR}
\end{equation}
This model possesses a considerable degree of generality and also, as a special case, it includes the standard linear recurrent model.

\subsection{Recursive Estimation of SSA with a General Recurrent Formula}
In this section, we are concerned with updating and estimation of the parameters,
\{$\phi_1,\phi_2,\ldots,\phi_{L-1}\}$  of model (\ref{SDMAR}) in the forecast period. However, these coefficients depend on the state vector  $Y_{t-1}$, and the estimation problem thus becomes the estimation of the functional form of this dependency. In order to estimate these coefficients, a recursive method similar to that of \cite{Harrison1976} is used.

Priestley \cite{Priestley1980} has shown it is possible to carry out the estimation procedure based on the extended Kalman Filter algorithm. However, there are also some assumptions about parameters \cite{Kalman}.  The simplest non-trivial assumption which can be made is that the  parameters are represented locally as linear functions of the state-vector  $Y_t$. Provided $\{\phi_u\}$, are slowly changing functions of  $Y_t$,  this assumption is valid.  Using these assumptions, ‘updating’ equations for the parameters may be written as:
\begin{align}
\phi_{u,t}&= \phi_{u,t-1}+\Delta{y_{t-u}} \gamma^{(t)}_u, \qquad u=1,\ldots,L-1, 
\end{align}
where $\Delta{y_{t-u}}=y_{t-u}-y_{t-u-1}$ and $\gamma_u$ is the gradient.\footnote{In general this may be defined as $\Delta{y_{t-u}}=y_{t-u}-y_{t-u-d}$ . For example, for monthly time series when $d=1$,  $\Delta{y_{t-u}}$  is considered as monthly difference and when $d=12$ it is considered as yearly difference.}
The ‘gradient’ parameters  $\gamma^{(t)}_1,\ldots,\gamma^{(t)}_{L-1}$ are unknowns, and must be estimated. The  basic strategy is to  allow these parameters to wander in the form of ‘random walks’. The random walk model for the gradient parameters may be written  in matrix form as:
\begin{equation}
{B}_{t+1}={B}_{t}+{V}_{t+1},
\end{equation}
while ${B}_t=(\gamma^{(t)}_1,\ldots,\gamma^{(t)}_{L-1})$ and ${V}_t$ is a sequence of independent matrix-valued random variables such that ${V}_t \sim N(0,{\bf\Sigma}_V)$. The estimation procedure then determines, for each $t$, those values of ${B}_t$, which roughly speaking, minimise the discrepancy between the observed value of  $y_{t+1}$ and its predictor $\hat{y}_{t+1}$, computed from the model fitted at time $t$.  The algorithm is thus sequential in nature and resembles the procedures used in the Kalman filter algorithm \cite{Kalman}. Following \cite{Priestley1980} and with some modifications, we can re-write the general recurrent model in a state-space form in which the state-vector is no longer $Y_t$, but is replaced by the state-vector:
\begin{equation}
{\theta}_{(t)}= (\phi^{(t-1)}_1,\ldots,\phi^{(t-1)}_{L-1},\gamma^{(t)}_1,\ldots,\gamma^{(t)}_{L-1}),
\end{equation}
i.e., $\theta_t$ is the vector of all current parameters of the model.  Applying the Kalman algorithm to the reformulated equations yields the recursion
\begin{equation}\label{hattheta}
\hat{\theta}_t=\mathbf{F}^*_{t-1}\theta_{t-1}+\mathbf{K}^*_t(Y_t-H_t^* \mathbf{F}^*_{t-1}\theta_{t-1}),
\end{equation}
\begin{equation}\label{Xt}
Y_t=H_t^* \theta_t+\epsilon_t,
\end{equation}
where $H^*_t=(Y_{t-1},\ldots,Y_{t-(L-1)},0,\ldots,0)$
       \[\mathbf{F}^*_{t-1}=\left(\begin{array}{c|ccc}
   \mathbf{ I}_{ L-1} & \begin{array}{ccccc}
   \Delta{y}_{t-1}&&&0\\
    & \Delta{y}_{t-2} \\
    & & \ddots\\
    0& & &  \Delta{y}_{t-(L-1)}
 \end{array} \\ \hline
    0  & \mathbf{I}_{L-1} \\
  \end{array}\right)
\]
where $\Delta{y}_{t-u}=(y_{t-u}-y_{t-u-1}); u=1,\ldots,L-1,$ and $ \mathbf{K}^*_{t}$, the ‘Kalman gain’ matrix, is given by:
\begin{equation}
\mathbf{K}^*_{t}=\mathbf{\Phi}_t(H^*_{t})^T \sigma^2_e,
\end{equation}
$\mathbf{\Phi}_t$ being the variance-covariance matrix of the one-step prediction error of $\theta_t$, i.e.
\begin{equation*}
\mathbf{\Phi}_t= E\Big[(\theta_t-\mathbf{F}^*_{t-1}\hat{\theta}_{t-1})(\theta_t-\mathbf{F}^*_{t-1}\hat{\theta}_{t-1})^T\Big],
\end{equation*}
and $\sigma^2_e$ is the variance of the one-step ahead prediction error of $y_t$, i.e., $\sigma^2_e$  is  the variance of  $e_t ={y_t -H^*_t \mathbf{F}^*_{t-1} \hat{\theta}_{t-1} }$.  If $\mathbf{C}_t$ is the variance-covariance matrix of $(\theta_t-\hat{ \theta}_t )$, then successive values of $\hat{\theta}_t$ may be estimated by using the standard recursive equations for the Kalman Filter as:
\begin{align}
\mathbf{\Phi}_t&= \mathbf{F}^*_{t-1}\mathbf{C}_{t-1}(\mathbf{F}^*_{t-1})^T+\mathbf{\Sigma}_W, \label{Phi}
\\[1em]
\mathbf{K}^*_{t}&=\mathbf{\Phi}_t(H^*_{t})^T\Big[ H^*_{t}\mathbf{\Phi}_t(H^*_{t})^T+\sigma^2_\epsilon \Big]^{-1} \label{K},
\\[1em]
\mathbf{C}_{t}&=\mathbf{\Phi}_t-\mathbf{K}^*_{t}\Big[ H^*_{t}\mathbf{\Phi}_t(H^*_{t})^T+\sigma^2_\epsilon \Big](\mathbf{K}^*_{t})^T, \label{Ct}
\end{align}
where
\begin{equation}
\mathbf{\Sigma}_W={\begin{pmatrix} 0&0\\ 0&{\mathbf{\Sigma}_V} \\ \end{pmatrix}}.
\end{equation}
Using the parameters estimated via the SSA (bootstrap SSA) technique with window length $L$, this recursive procedure will be started at the beginning of the forecast period, with the initial estimated coefficients as:
  $\hat{\theta}_{t_0-1}= ({\hat{\phi}_1},\ldots,{\hat{\phi}_{L-1}},0,\ldots,0)$ and the residual variance of the model ${\hat{\mathbf{\sigma}}^2_{\epsilon}}$ and
\begin{equation}
\mathbf{C}_{t_0-1}={\begin{pmatrix} \hat{\mathbf{R}}_{\phi_1,\ldots,\phi_{L-1}}&0\\ 0&0 \\ \end{pmatrix}}.
\end{equation}
If it is assumed that the values of the parameters obtained within the sample period for the linear recurrent formula are optimal, it also seems reasonable to set  all the gradients to zero initially.  It remains to choose reasonable values for $\mathbf{\Sigma}_V$, the variance-covariance matrix of ${V}_{t}$ (and hence, by implication, to choose values for $\mathbf{\Sigma}_W$). The choice of $\mathbf{\Sigma}_V$ depends on the assumed ‘smoothness’ of the recurrent parameters within the forecast period as functions of $Y_t$. The diagonal elements of $\mathbf{\Sigma}_V$  are set equal to $\sigma^2_{\epsilon}$ multiplied by some constant  called the ‘smoothing factor’, and the off-diagonal elements are set equal to zero.  However, if the elements of $\mathbf{\Sigma}_V$ are set too small, it is difficult to detect the structural breaks or non-linearity present in the data and the parameters stay the same as those obtained within the sample period by SSA and almost constant in the forecast period.

As mentioned before, the gradient parameters $\gamma^{(t)}_1,\ldots,\gamma^{(t)}_{L-1}$ are treated as random walks with respect to time, the variance of the innovation term in the random walk being flexible and determined by the smoothing factor chosen by the user. As suggested in \cite{Haggan1984}, we also  select the smoothing factor in the range of $10^{-3}$  to $10^{-6}$ in this study.

\section{Descriptive Statistics and Structural Breaks of the Data}
The data used in this study are taken from  I.N.S.E.E (Institute National de la Statistiuqe et des Etudes Economiques) for France, from Statistisches Bundesamt, Wesbaden for Germany and from the Office for National Statistics (ONS) for the UK and represents eight major components of real industrial production in France, Germany and the UK. The series are seasonally adjusted monthly indices for real output in electricity and gas, chemicals, fabricated metals, vehicle, food products, basic metals, electrical machinery and machinery.   The eight series examined  are interesting, important and reflect diverse types of industries, ranging from traditional industrial sectors such as machinery and basic metals to electricity/gas and food products. Although we consider only eight of the two digit industries in this study, these eight industries account for more than 50$\%$ of the total industrial production in each country. The same  eight  industries have been considered by \cite{Heravi2004}, \cite{Hassani2009} and \cite{Osborn1999}.

\begin{figure}
        \centering
        \begin{subfigure}[h]{0.45\textwidth}
                \includegraphics[width=1\textwidth,height=1.7in]{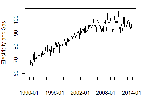}
                \label{Electricity and Gas}
        \end{subfigure}%
        \begin{subfigure}[h]{0.45\textwidth}
                \includegraphics[width=1\textwidth,height=1.7in]{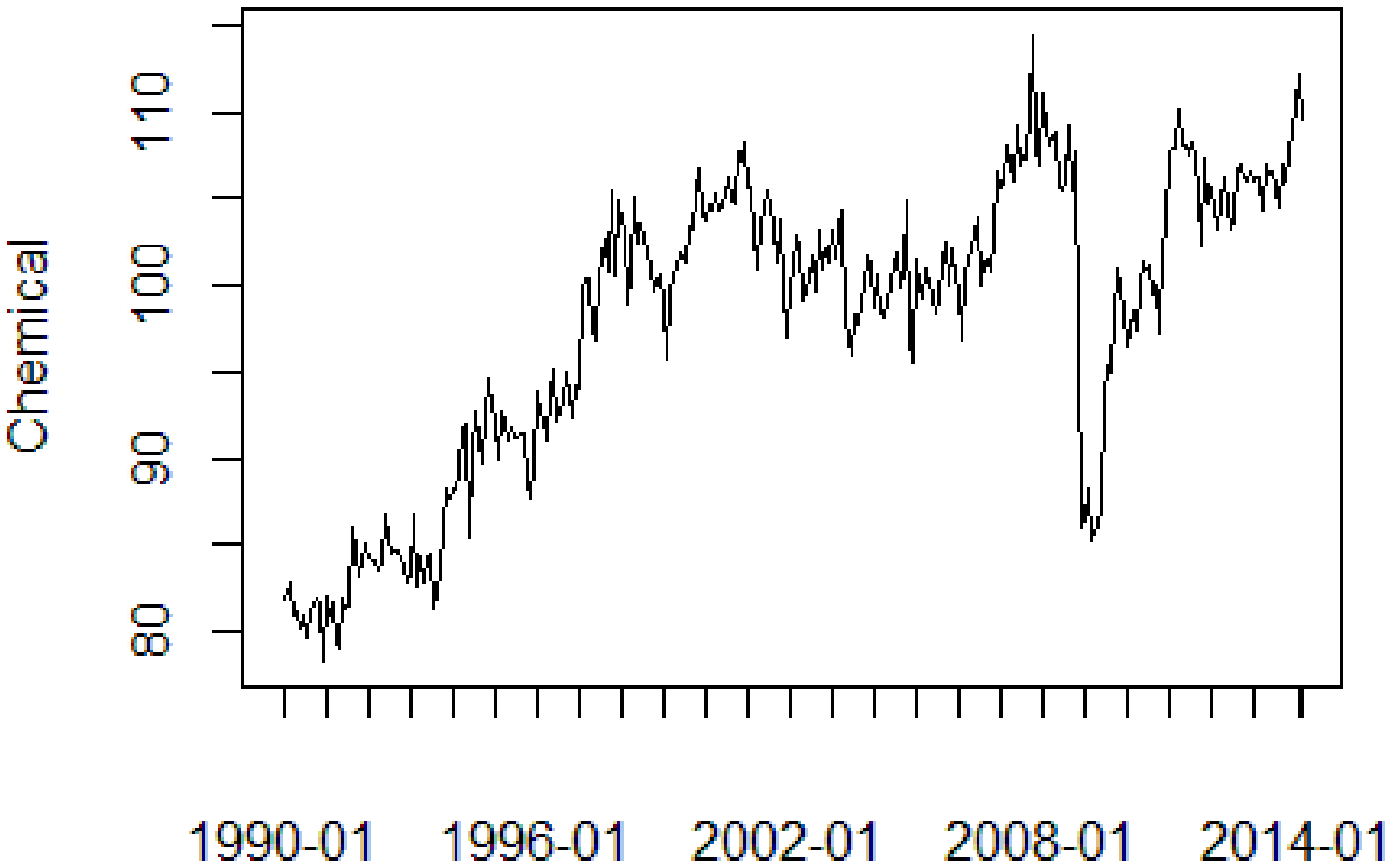}
                \label{Chemical}
        \end{subfigure}
       \begin{subfigure}[h]{0.45\textwidth}
                \includegraphics[width=1\textwidth,height=1.7in]{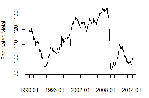}
                \label{Fabricate Metals}
        \end{subfigure}
        \begin{subfigure}[h]{0.45\textwidth}
                \includegraphics[width=1\textwidth,height=1.7in]{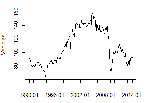}
                \label{Vehicle}
        \end{subfigure}
        \begin{subfigure}[h]{0.45\textwidth}
                \includegraphics[width=1\textwidth,height=1.7in]{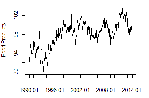}
                \label{Food}
        \end{subfigure}
        \begin{subfigure}[h]{0.45\textwidth}
                \includegraphics[width=1\textwidth,height=1.7in]{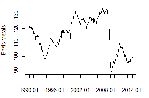}
                \label{Basic Metals}
        \end{subfigure}
        \begin{subfigure}[h]{0.45\textwidth}
                \includegraphics[width=1\textwidth,height=1.7in]{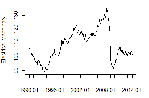}
        \end{subfigure}
        \begin{subfigure}[h]{0.45\textwidth}
                \includegraphics[width=1\textwidth,height=1.7in]{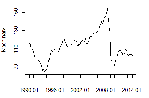}
                \label{Electrical Machinery}
        \end{subfigure}
\caption{Industrial Production Indicators for France}
\label{fig:France}
\end{figure}

\begin{figure}
        \centering
        \begin{subfigure}[h]{0.45\textwidth}
                \includegraphics[width=1\textwidth,height=1.7in]{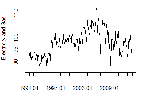}
                \label{Electricity and Gas}
        \end{subfigure}%
        \begin{subfigure}[h]{0.45\textwidth}
                \includegraphics[width=1\textwidth,height=1.7in]{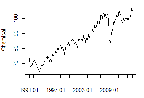}
                \label{Chemical}
        \end{subfigure}
       \begin{subfigure}[h]{0.45\textwidth}
                \includegraphics[width=1\textwidth,height=1.7in]{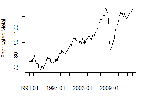}
                \label{Fabricate Metals}
        \end{subfigure}
        \begin{subfigure}[h]{0.45\textwidth}
                \includegraphics[width=1\textwidth,height=1.7in]{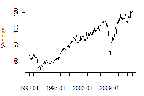}
                \label{Vehicle}
        \end{subfigure}
        \begin{subfigure}[h]{0.45\textwidth}
                \includegraphics[width=1\textwidth,height=1.7in]{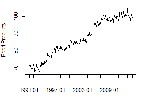}
                \label{Food}
        \end{subfigure}
        \begin{subfigure}[h]{0.45\textwidth}
                \includegraphics[width=1\textwidth,height=1.7in]{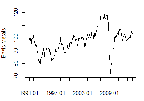}
                \label{Basic Metals}
        \end{subfigure}
        \begin{subfigure}[h]{0.45\textwidth}
                \includegraphics[width=1\textwidth,height=1.7in]{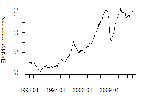}
                \label{Machinery}
        \end{subfigure}
        \begin{subfigure}[h]{0.45\textwidth}
                \includegraphics[width=1\textwidth,height=1.7in]{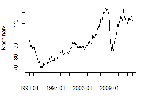}
                \label{Electrical Machinery}
        \end{subfigure}
\caption{Industrial Production Indicators for Germany }
\label{fig:Germany}
\end{figure}

\begin{figure}
        \centering
        \begin{subfigure}[h]{0.45\textwidth}
                \includegraphics[width=1\textwidth,height=1.7in]{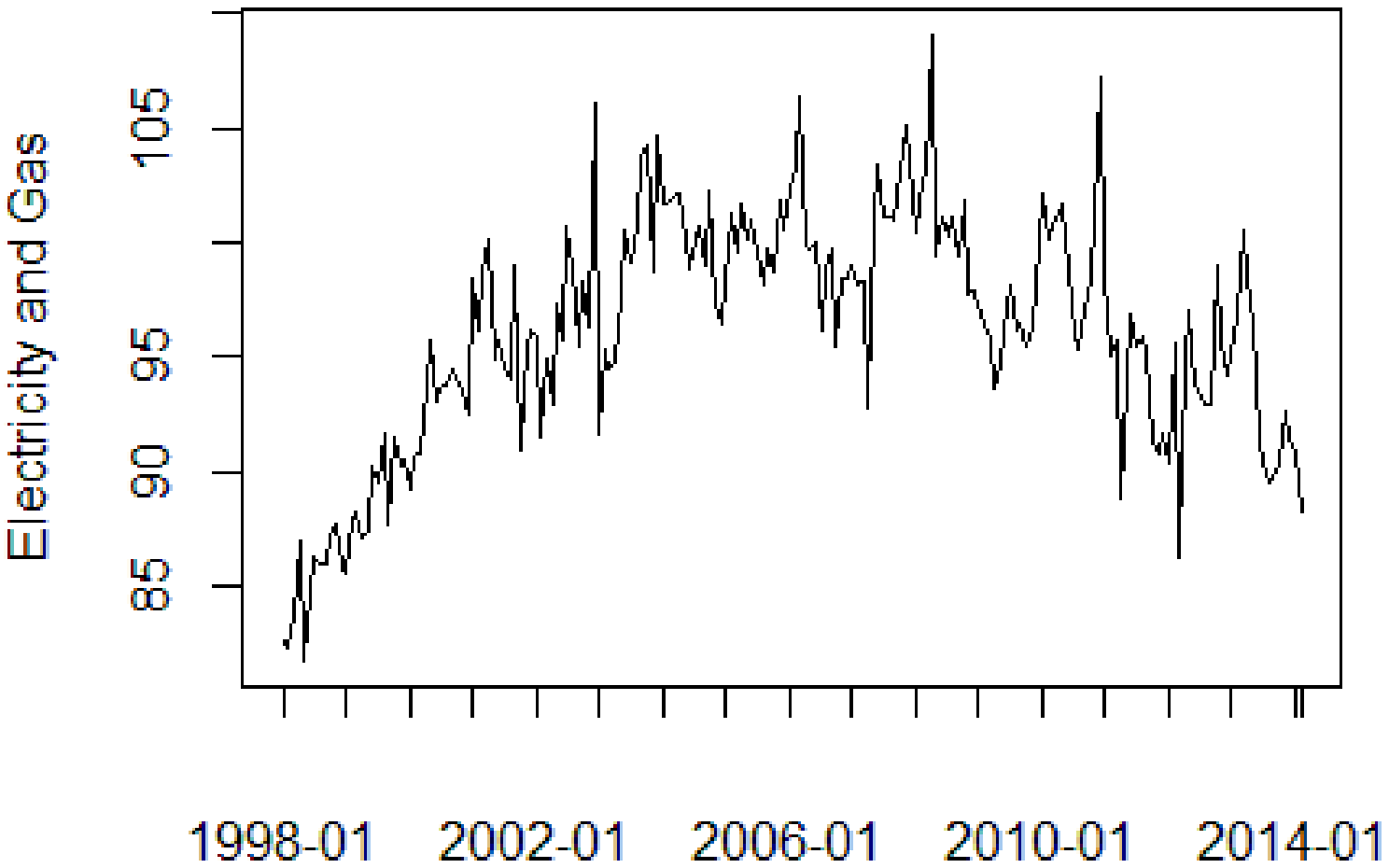}
                \label{Electricity and Gas}
        \end{subfigure}%
        \begin{subfigure}[h]{0.45\textwidth}
                \includegraphics[width=1\textwidth,height=1.7in]{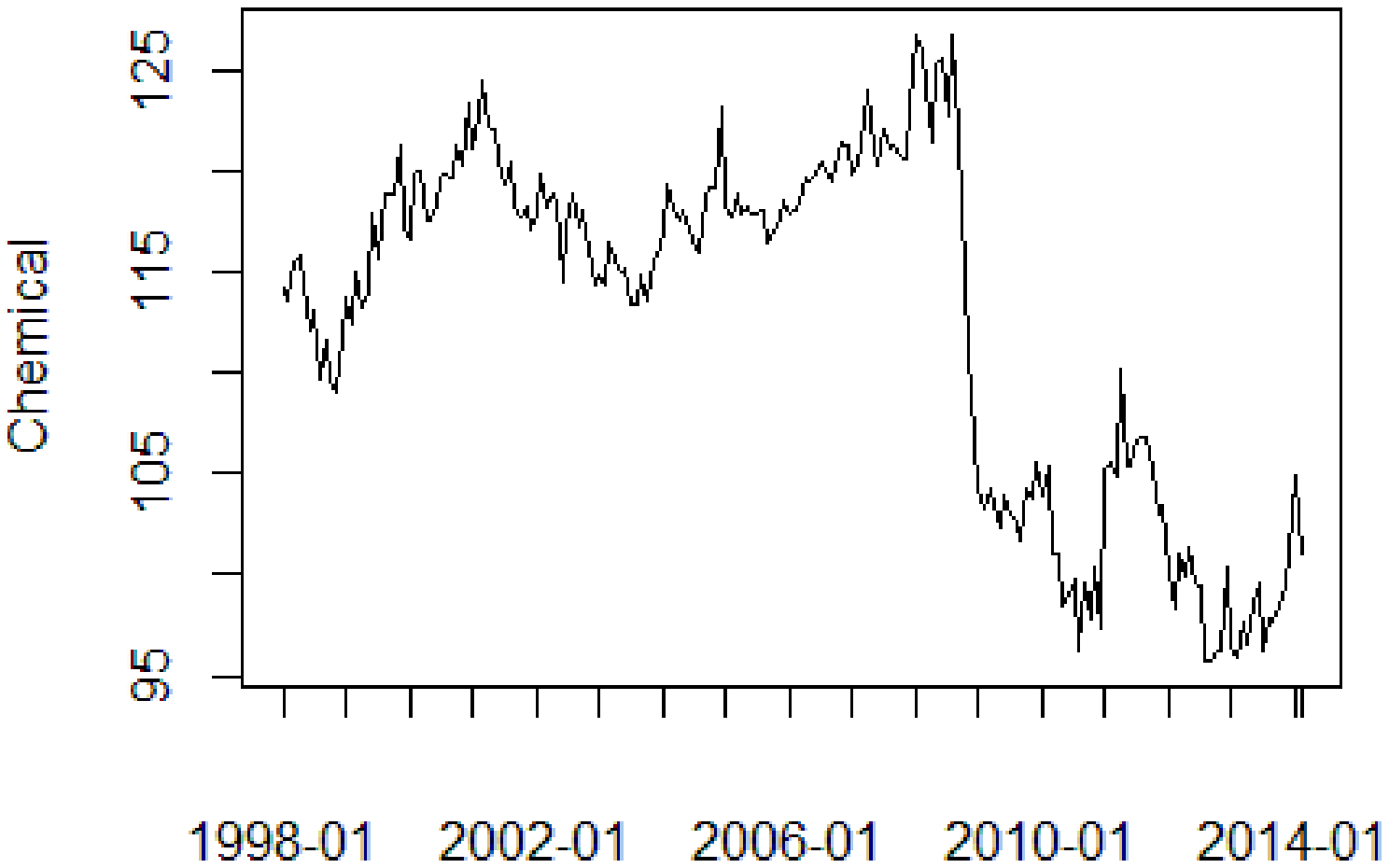}
                \label{Chemical}
        \end{subfigure}
       \begin{subfigure}[h]{0.45\textwidth}
                \includegraphics[width=1\textwidth,height=1.7in]{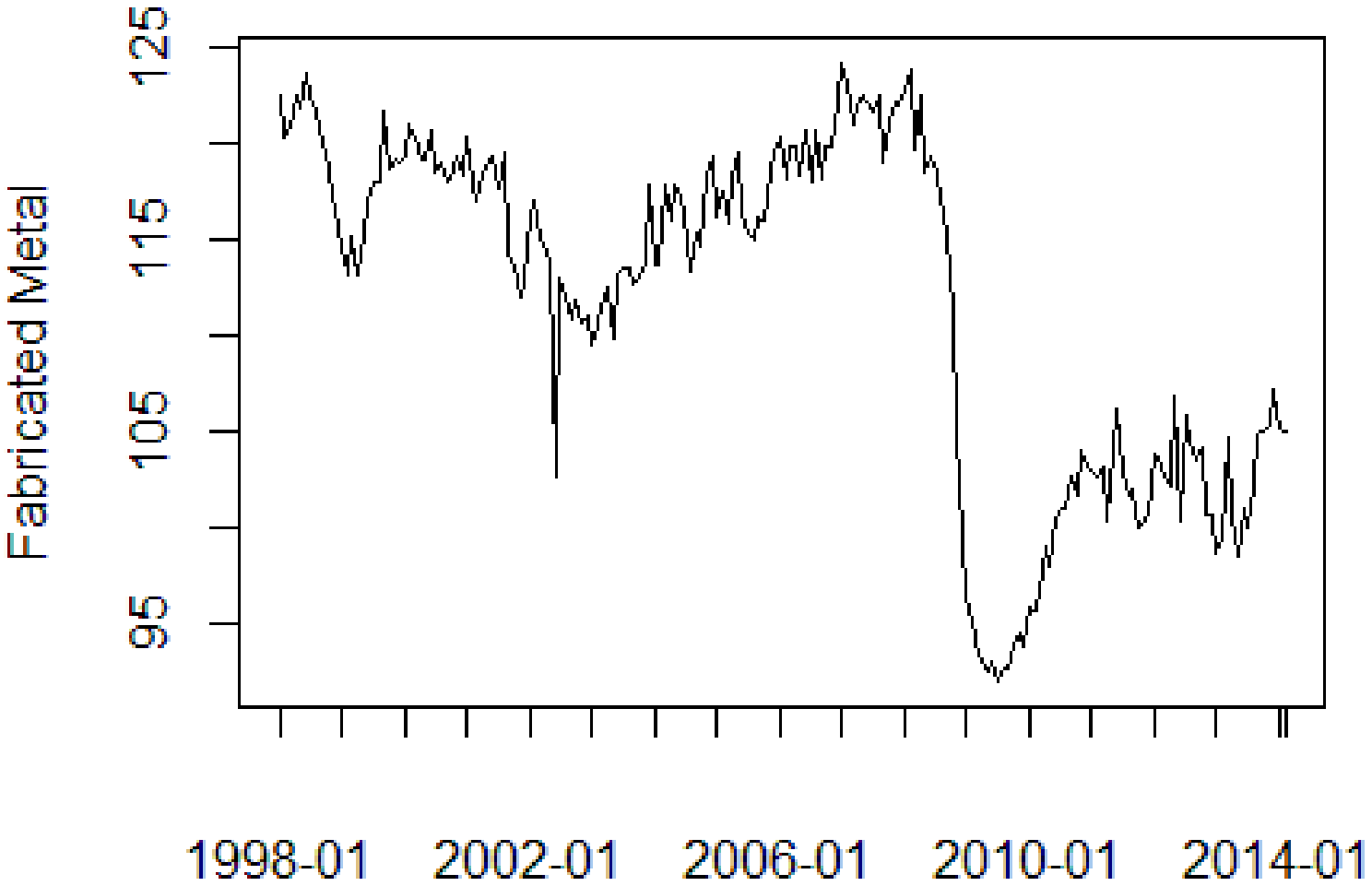}
                \label{Fabricate Metals}
        \end{subfigure}
        \begin{subfigure}[h]{0.45\textwidth}
                \includegraphics[width=1\textwidth,height=1.7in]{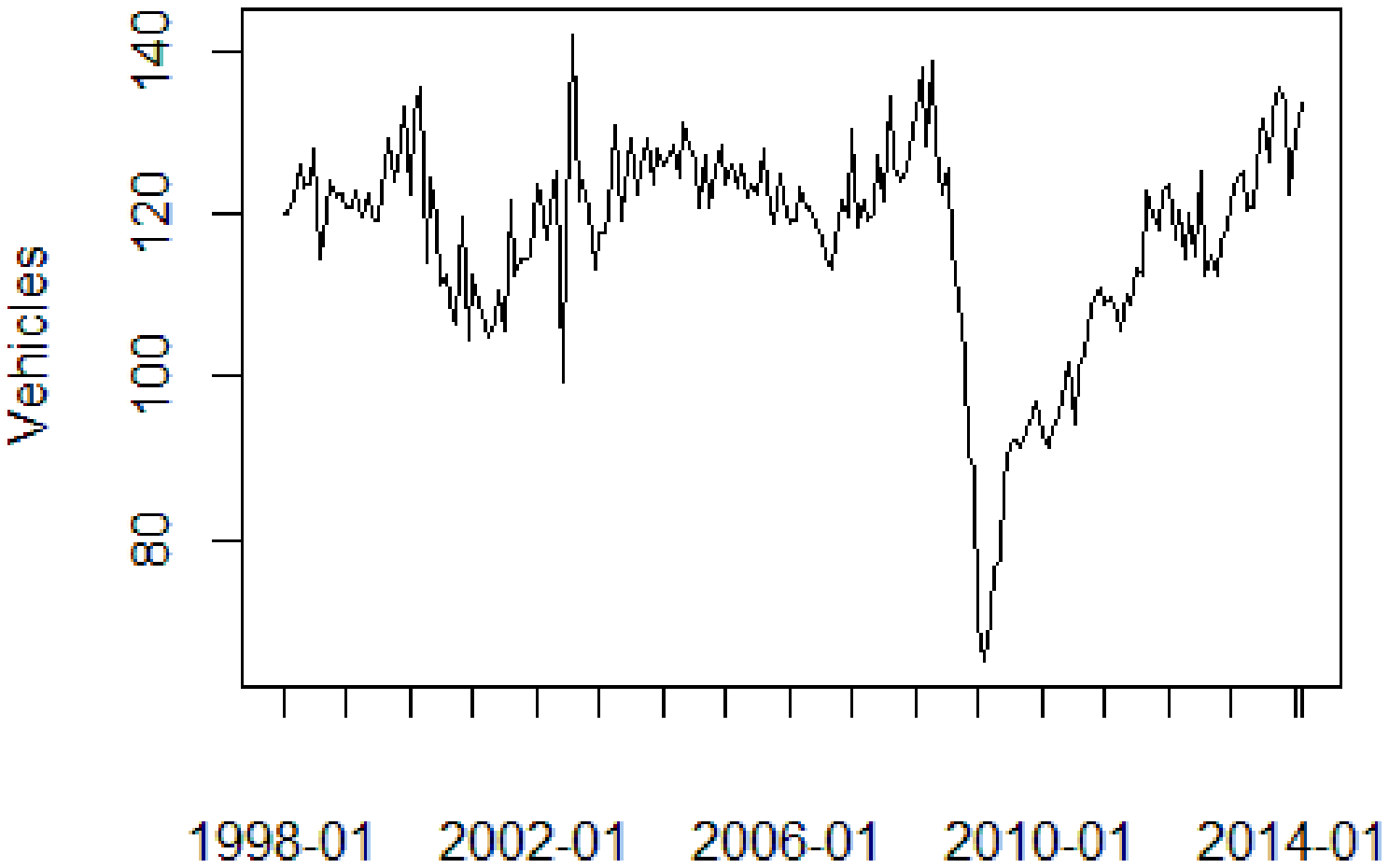}
                \label{Vehicle}
        \end{subfigure}
        \begin{subfigure}[h]{0.45\textwidth}
                \includegraphics[width=1\textwidth,height=1.7in]{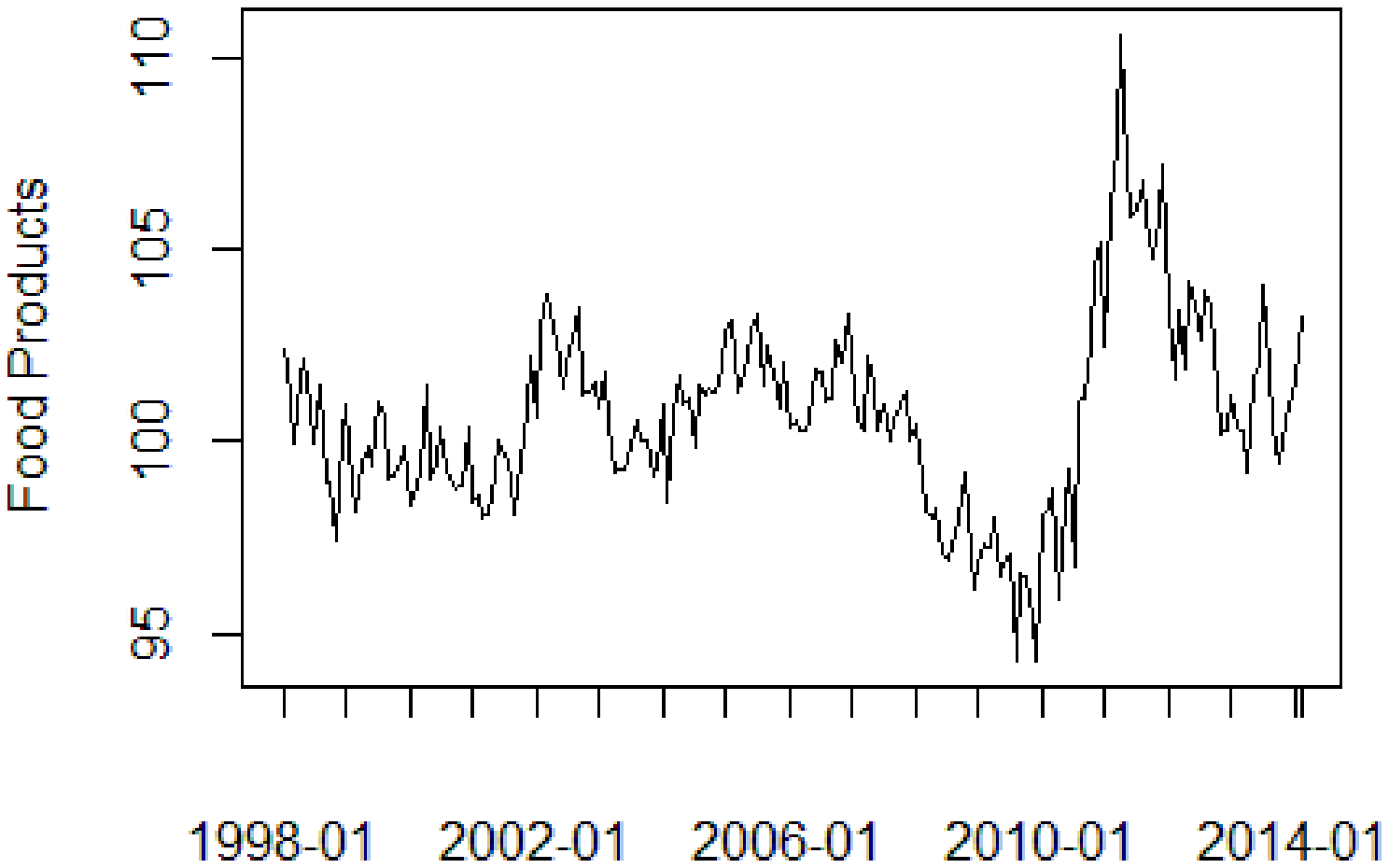}
                \label{Food}
        \end{subfigure}
        \begin{subfigure}[h]{0.45\textwidth}
                \includegraphics[width=1\textwidth,height=1.7in]{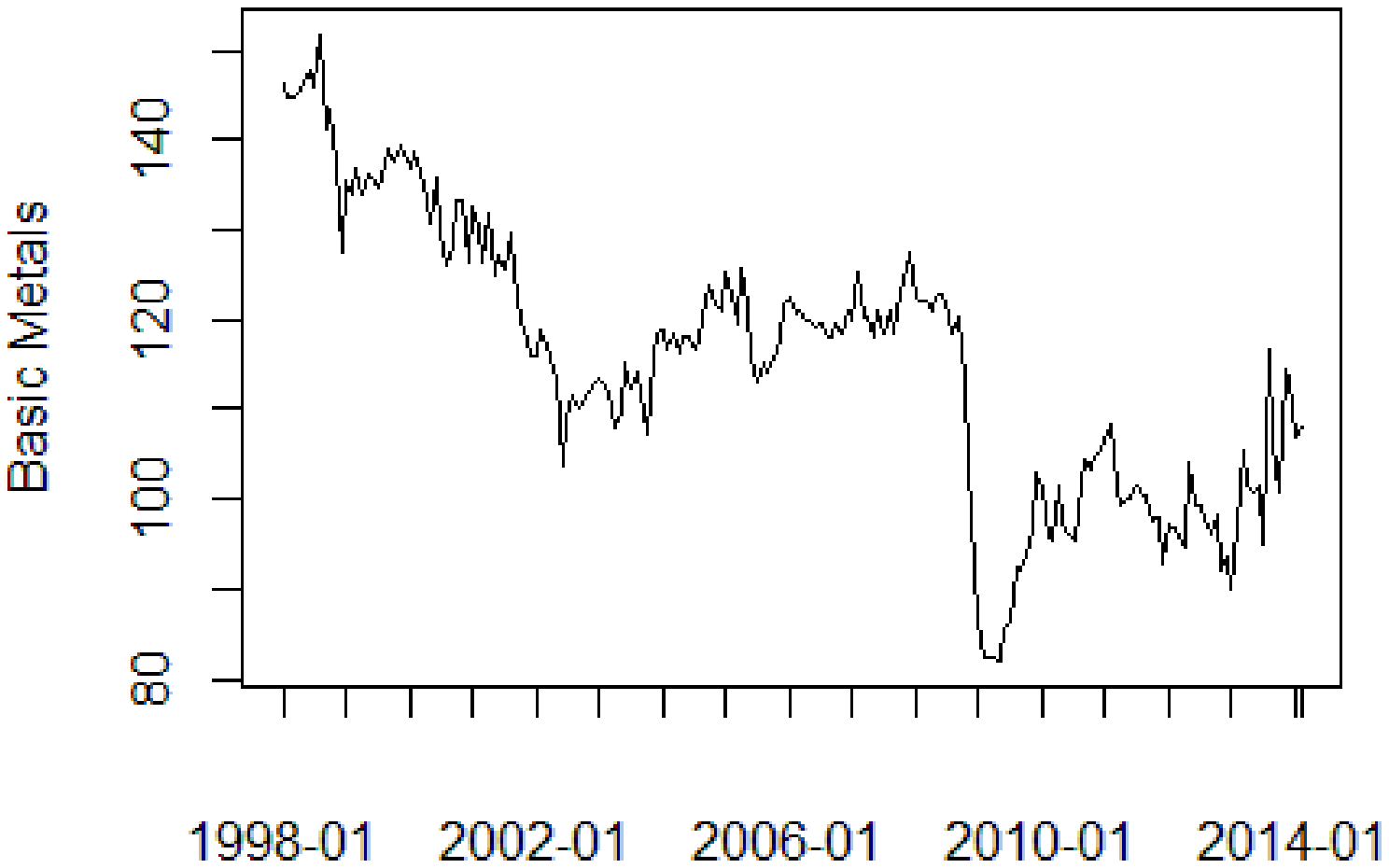}
                \label{Basic Metals}
        \end{subfigure}
        \begin{subfigure}[h]{0.45\textwidth}
                \includegraphics[width=1\textwidth,height=1.7in]{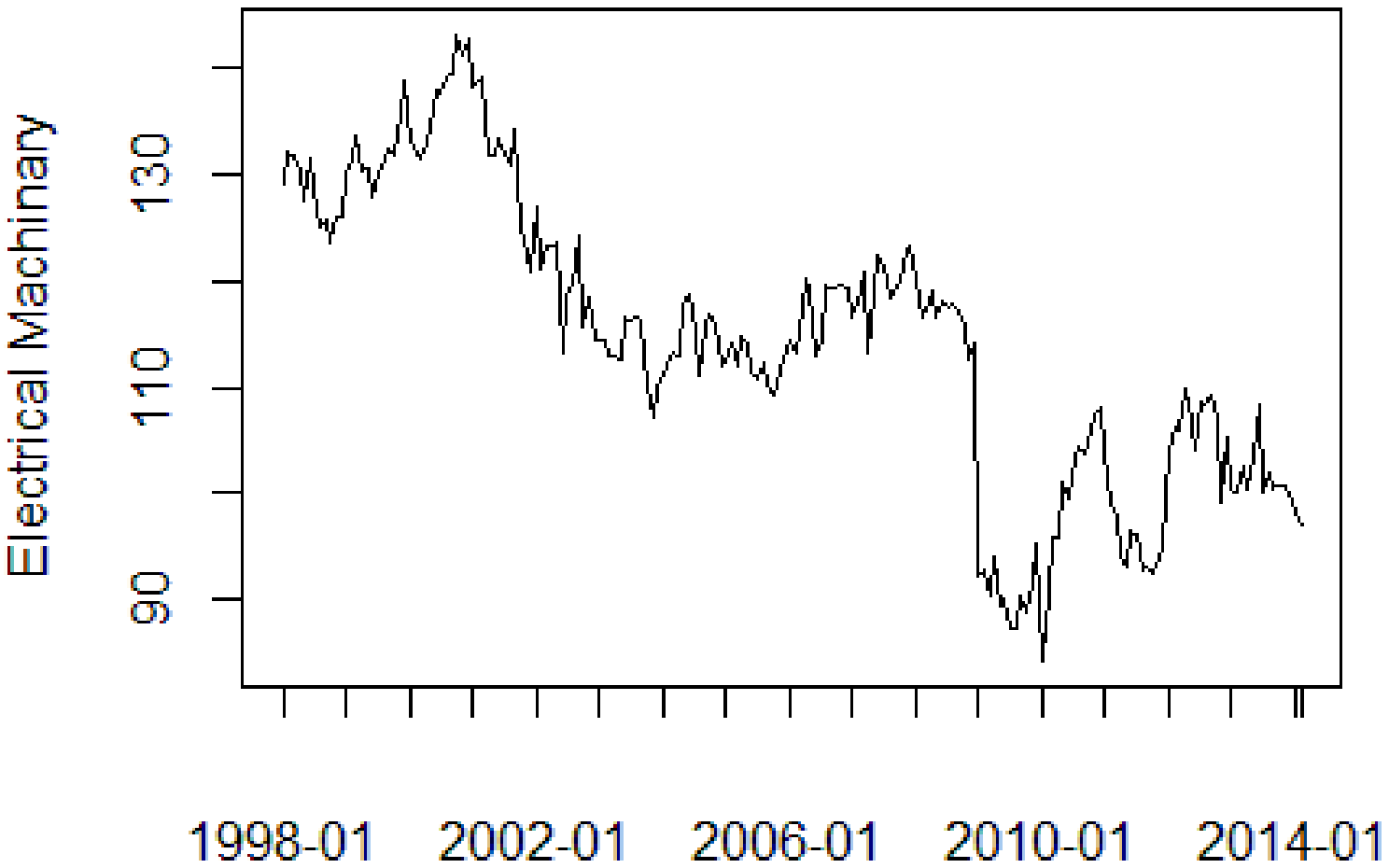}
                \label{Machinery}
        \end{subfigure}
        \begin{subfigure}[h]{0.45\textwidth}
                \includegraphics[width=1\textwidth,height=1.7in]{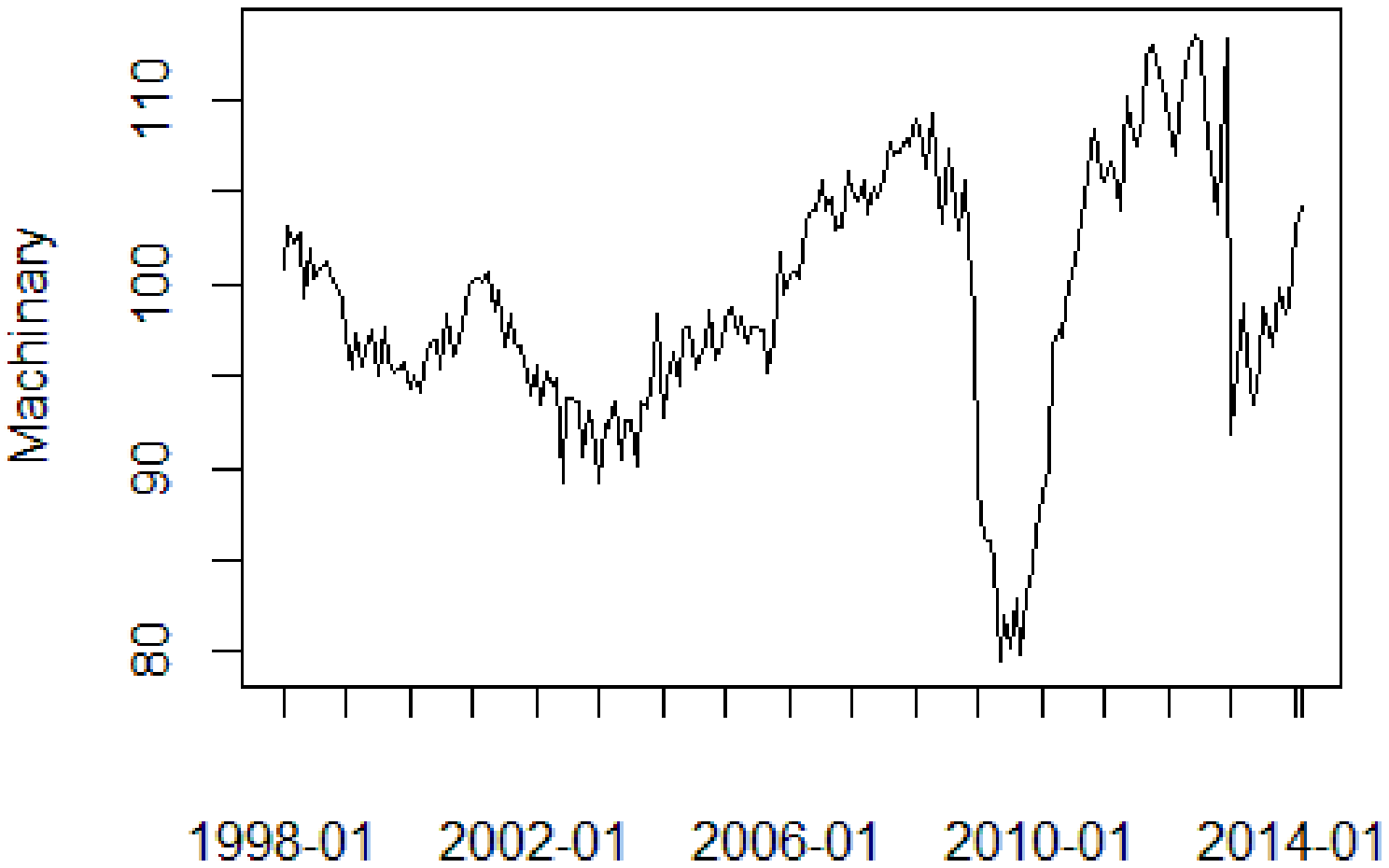}
                \label{Electrical Machinery}
        \end{subfigure}
\caption{Industrial Production Indicators for UK}
\label{fig:UK}
\end{figure}

In all cases our sample period ends in February 2014. However, the data for France  start from January 1990, for Germany from January 1991 and for the UK start from January 1998.  Figures \ref{fig:France}, \ref{fig:Germany} and \ref{fig:UK} show the series used in this study. Periods of overall expansion and contraction are evident in the graphs. As can be seen, most series for France and Germany present a long period of growth in 1990s and up to the current recession of 2008-2009. For the UK, however, most series show a period of stagnation in early 2000s and recession in 2008. For almost all the series, the steep drop in production can be seen around 2008-2009, which is attributed to the banking crisis and current recession.

Tables \ref{fr1},\ref{ger1} and \ref{uk1} also show the Bai and Perron test \cite{Bai} for finding the structural breaks in these data series. Almost all the series show a break or multiple breaks in the data period considered. However, only the date of the last break point is reported for each series in the tables to save space (Full detailed of structural break points are available from the authors upon request). Results for all the three countries, based on the Bai and Perron test, indicate that all sectors are affected by the current recession of 2008-2009, except for the electricity and gas for Germany. The results also indicate the break points for food and chemicals in June and July 2010 for Germany and May 2010 for food in France.

Tables \ref{fr1},\ref{ger1} and \ref{uk1}  also show the descriptive statistics for the monthly percentage changes in the original series, i.e. $100((y_t-y_{t-1})/y_{t-1})$. In addition to reporting the  growth/decline for the whole period, we also report the monthly percentage changes before and after the break points for each series. Overall all sectors have experienced growth over the whole period, with the exception of basic metals, fabricated metals and machinery for France and the UK. In addition, for the UK, chemicals also shows a decline over this period. Some sectors have experienced  substantial growth in production over 1990s and early 2000s for Germany and France. In particular, electricity/gas and vehicle in France show average increase of around 0.30 percentage  per month or 3.6$\%$ per year. The results for Germany show that all sectors have recovered  after the recession, with the exception of food products, and in particular vehicles shows an average increase of 0.64 percentage  or about 8$\%$ growth per year. Declining industries after the break points are mostly in France and the UK, with machinery showing the highest average decline of 0.41$\%$  per month.
The sample standard deviations indicate greater volatility  for the vehicle series than those of other sectors and with very low volatility for food products. The results for normality test based on Shapiro-Wilk test also provide strong evidence of non-normality for all the series, except for the food products. The results are all statistically significant at 1$\%$ level except electricity/gas for Germany and food products for all the three countries.

\begin{minipage}[t]{.9\textwidth}
 \begin{center}
 \captionof{table}{Descriptive statistics of Industrial Production Indicators for France. }\label{fr1}
 \begin{adjustbox}{center, width=\columnwidth-60pt}
\begin{tabular}{cccccccccccc}
\toprule
&Weight&\vtop{\hbox{\strut Sample}\hbox{\strut Size}}&\vtop{\hbox{\strut In-sample} \hbox{\strut Mean}}&\vtop{\hbox{\strut Out-Of-Sample} \hbox{\strut Mean}}&\vtop{\hbox{\strut Overall}\hbox{\strut Mean}}&SD&SW(p)&\vtop{\hbox{\strut Final}\hbox{\strut Break Points}}&Out-Of-Sample\\
  \midrule
Electricity and gas	&	9.00	&	290	&	0.29	&	0.04	&	0.22	&	4.46	&	0.00*	&	May-09	&	69 \\
Chemical	&	8.90	&	290	&	0.09	&	0.42	&	0.13	&	2.46	&	0.00*	&	May-10	&	57	\\
Fabricate Metals	&	4.30	&	290	&	0.08	&	-0.34	&	-0.03	&	2.36	&	0.00*	&	Nov-08	&	75	\\
Vehicle	&	9.80	&	290	&	0.29	&	-0.27	&	0.13	&	5.06	&	0.00*	&	Apr-08	&	82	\\
Food Products	&	8.60	&	290	&	0.06	&	0.02	&	0.04	&	1.65	&	0.11	&	Jan-08	&	85	\\
Basic Metals	&	3.90	&	290	&	0.06	&	-0.35	&	-0.05	&	2.07	&	0.00*	&	Nov-08	&	75	\\
Electrical Machinery	&	7.10	&	290	&	0.10	&	-0.28	&	0.01	&	2.06	&	0.00*	&	Nov-08	&	75	\\
Machinery	&	9.60	&	290	&	0.11	&	-0.41	&	-0.02	&	1.93	&	0.00*	&	Nov-08	&	75	\\
\hline
\end{tabular}
 \end{adjustbox}
  \end{center}
 \center
\scriptsize{\textit{Note}:* indicates results are statistically significant at p=0.01 based on \\ Shapiro-Wilk test.}
\end{minipage}
\newline
\newline
\begin{minipage}[H]{.9\textwidth}
    \begin{center}
     \captionof{table}{Descriptive statistics of Industrial Production Indicators for Germany.}\label{ger1}
        \begin{adjustbox}{center, width=\columnwidth-70pt}
\begin{tabular}{cccccccccccc}
\toprule
&Weight&\vtop{\hbox{\strut Sample}\hbox{\strut Size}}&\vtop{\hbox{\strut In-sample} \hbox{\strut Mean}}&\vtop{\hbox{\strut Out-Of-Sample} \hbox{\strut Mean}}&\vtop{\hbox{\strut Overall}\hbox{\strut Mean}}&SD&SW(p)&\vtop{\hbox{\strut Final}\hbox{\strut Break Points}}&Out-Of-Sample\\
  \midrule
Electricity and gas	&	7.60	&	278	&	0.06	&	0.01	&	0.04	&	2.83	&	0.02	&	Nov-02	&	123	\\
Chemical	&	8.60	&	278	&	0.15	&	0.09	&	0.16	&	2.18	&	0.00*	&	Jul-10	&	31	\\
Fabricate Metals	&	4.50	&	278	&	0.11	&	0.50	&	0.18	&	2.01	&	0.00*	&	Oct-08	&	52	\\
Vehicle	&	7.20	&	278	&	0.24	&	0.64	&	0.33	&	4.93	&	0.00*	&	Sep-08	&	53	\\
Food Products	&	13.60	&	278	&	0.16	&	-0.04	&	0.13	&	1.94	&	0.53	&	Jun-10	&	32	\\
Basic Metals	&	5.60	&	278	&	0.03	&	0.30	&	0.07	&	3.20	&	0.00*	&	Oct-08	&	52	\\
Electrical Machinery	&	10.40	&	278	&	0.20	&	0.43	&	0.24	&	2.07	&	0.00*	&	Dec-08	&	50	\\
Machinery	&	6.50	&	278	&	0.02	&	0.52	&	0.11	&	2.97	&	0.00*	&	Dec-08	&	50	\\
\hline
\end{tabular}
 \end{adjustbox}
    \end{center}
 \center
\scriptsize{\textit{Note}:* indicates results are statistically significant at p=0.01 based on \\ Shapiro-Wilk test.}
\end{minipage}
\newline
\newline
\begin{minipage}[t]{.9\textwidth}
    \begin{center}
     \captionof{table}{Descriptive statistics of Industrial Production Indicators for  UK.}\label{uk1}
        \begin{adjustbox}{center, width=\columnwidth-70pt}
\begin{tabular}{cccccccccccc}
\toprule
&Weight&\vtop{\hbox{\strut Sample}\hbox{\strut Size}}&\vtop{\hbox{\strut In-sample} \hbox{\strut Mean}}&\vtop{\hbox{\strut Out-Of-Sample} \hbox{\strut Mean}}&\vtop{\hbox{\strut Overall}\hbox{\strut Mean}}&SD&SW(p)&\vtop{\hbox{\strut Final}\hbox{\strut Break Points}}&Out-Of-Sample\\
  \midrule
Electricity and gas	&	10.20	&	194	&	0.18	&	-0.17	&	0.09	&	3.25	&	0.00*	&	Oct-08	&	64	\\
Chemical	&	8.50	&	194	&	-0.01	&	-0.17	&	-0.05	&	1.91	&	0.01*	&	Oct-08	&	64	\\
Fabricate Metals	&	3.80	&	194	&	-0.10	&	-0.06	&	-0.06	&	2.12	&	0.00*	&	Oct-08	&	64	\\
Vehicle	&	5.80	&	194	&	0.10	&	0.37	&	0.22	&	5.72	&	0.00*	&	Aug-08	&	66	\\
Food Products	&	7.50	&	194	&	-0.02	&	0.05	&	0.01	&	1.36	&	0.60	&	Jan-08	&	73	\\
Basic Metals	&	3.00	&	194	&	-0.22	&	0.06	&	-0.08	&	3.86	&	0.00*	&	Oct-09	&	64	\\
Electrical Machinery	&	4.70	&	194	&	-0.21	&	-0.16	&	-0.09	&	3.22	&	0.00**	&	Dec-09	&	62	\\
Machinery	&	6.70	&	194	&	0.07	&	0.00	&	0.06	&	2.73	&	0.00*	&	Apr-09	&	70	\\
\hline
\end{tabular}
 \end{adjustbox}
    \end{center}
 \center
\scriptsize{\textit{Note}:* indicates results are statistically significant at p=0.01 based on \\ Shapiro-Wilk test.}
\end{minipage}
\newline
\section{Forecasting results}
We now turn to an issue of central interest in this paper, namely the evaluation of forecast performance of the Singular Spectrum Analysis  with the “general linear recurrent” model (GSSA) with the bootstrap SSA and basic SSA with the “linear recurrent” formula. In addition, we assess the forecasting accuracy for  four different horizons, 1$-$step ahead, 3 and 6$-$steps ahead and one year ahead. All models are estimated  based on the data up to the final break points given in tables \ref{fr1},\ref{ger1} and \ref{uk1}, as our interest is to assess the forecast accuracy in the presence of a structural break in the forecast period. Post-sample forecasts are then computed for the months from the final break point to the end of the data, February 2014. Thus the number of observations  retained for post-sample forecast accuracy test are different depending on the date of the break point in the series. However, as may be seen form the descriptive tables in section 4, the number of observations retained for post-sample forecast accuracy evaluation are around 60 months, the minimum number of observations hold are 31 and 32 months for chemicals and food products for Germany.

Forecast accuracy is measured based on the magnitude of forecast errors, such as  the Root Mean Square Error (RMSE) and Mean Absolute Error (MAE). However, since these measures give quantitatively similar results and to conserve space, we only report the RMSE, as this is the most frequently quoted measure in forecasting \cite{Zhang1998}.  Tables \ref{fr2},\ref{ger2} and \ref{uk2} show the out-of-sample RMSE and the ratio of  RMSE (RRMSE) results for France, Germany and the UK. The ratio of RMSE here defined as

\begin{equation}
RRMSE=\left({\sum_{i=1}^{N}(\hat{y}_{T+h,i}-{y}_{y+h,i})^2}\right)^{\frac{1}{2}}/\left({\sum_{i=1}^{N}(\tilde{y}_{T+h,i}-{y}_{T+h,i})^2}\right)^{\frac{1}{2}}
\end{equation}

Where $\hat{y}_{T+h,i}$ is the $h$-step ahead forecast obtained by SSA forecasting and $\tilde{y}_{T+h,i}$ is the $h$-step ahead forecast from either Bootstrap SSA or General SSA  and $N$ is the number of forecasts. If RRMSE $<1.0$  then the general SSA outperform the SSA. Using the modified Diebold-Mariano statistics, given in \cite{Harvey1997}, we also test for the statistical significance of the results. Average RRMSE  is also given for each horizon at the bottom of each table.

In order to obtain the average bootstrap forecasts we have replicated the procedure 1000 times. The results show no evidence of any statistical difference between the SSA and the Bootstrap SSA, and in fact, they are very similar for all the horizons and all the three countries.  Comparing the GSSA  with the SSA, the results are significant for almost all the horizons and all the three countries. However, the quality of the forecast with general SSA is much better for $h=$1,3 and 6 and less significant for $h=$12. The general SSA technique outperforms SSA and reduces the RMSE by 40$\%$ for France and Germany and 28$\%$ for the UK. The improvements for $h=$12 are 10$\%$ for France and Germany and 18$\%$ for the UK.

\begin{figure}[H]
        \centering
 \begin{subfigure}[t]{0.5\textwidth}
                \includegraphics[width=1\textwidth,height=3in]{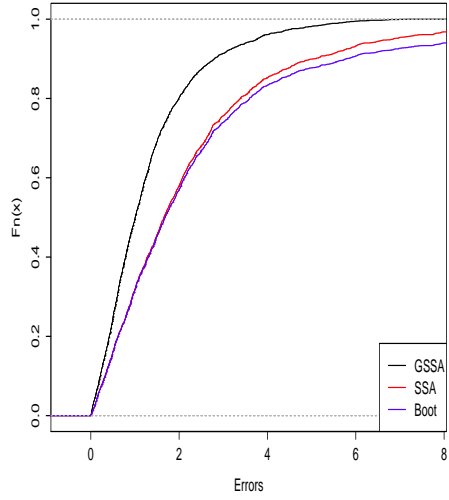}
                \caption{One-step-ahead}
                \label{ecdf1}
        \end{subfigure}%
        \begin{subfigure}[t]{0.5\textwidth}
                \includegraphics[width=1\textwidth,height=3in]{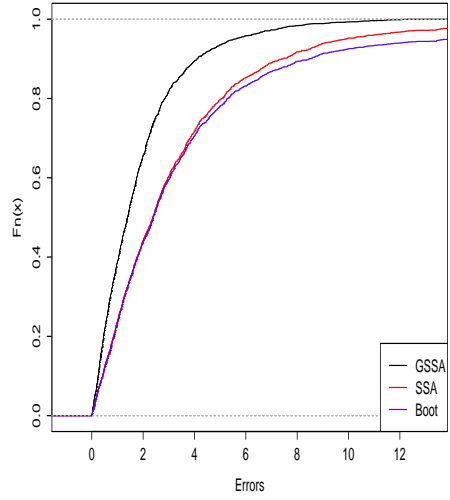}
                \caption{Three-step-ahead}
                \label{ecdf3}
        \end{subfigure}%

        \begin{subfigure}[t]{0.5\textwidth}
                \includegraphics[width=1\textwidth,height=3in]{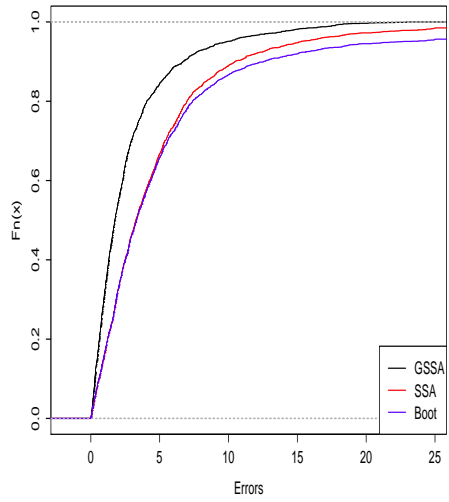}
                \caption{Six-step-ahead}
                \label{ecdf6}
        \end{subfigure}%
        \begin{subfigure}[t]{0.5\textwidth}
                \includegraphics[width=1\textwidth,height=3in]{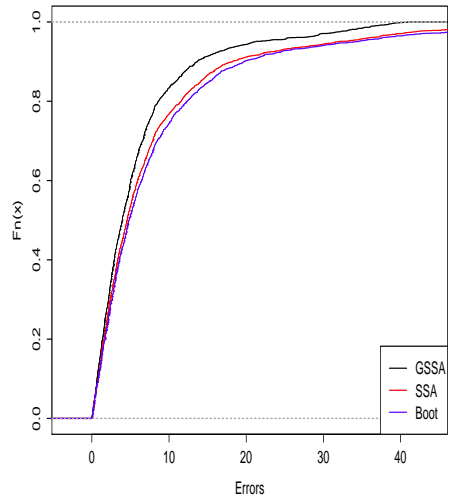}
                \caption{Twelve-step-ahead}
                \label{ecdf12}
        \end{subfigure}
 \caption[]{Empirical cumulative distribution functions of absolute values of forecast errors for SSA, Bootstrap SSA and General SSA.}
\label{ecdf}
\end{figure}

Figure \ref{ecdf} presents the cumulative distribution function (c.d.f) of the RMSE values of the absolute values of the out-of-sample errors obtained by the SSA, Bootstrap SSA and the general SSA for all the 24 series. If the c.d.f. produced by one method is strictly above the c.d.f. obtained by another method, we may then say that the forecast errors are stochastically smaller for the first method. Figures \ref{ecdf1},\ref{ecdf3},\ref{ecdf6} and \ref{ecdf12} demonstrate that the forecast errors obtained by the general SSA are much smaller than the errors of the other two methods for $h=$1,3, 6 and 12, confirming the superiority of the general SSA.

We also compute and report the percentage of forecasts that correctly predict the direction of change. Here the direction of change is interpreted in terms of monthly growth or decline in production over one month period in a particular sector. Tables \ref{fr2},\ref{ger2} and \ref{uk2} provide the percentage of forecasts that correctly predict the direction of change at h$=$1,3,6 and 12 months. At the bottom of the table, average for all the series is given for $h=$1,3,6 and 12 for each country. The percentage of the correct signs are not much higher than 50$\%$ and there is no evident prevalence of any particular forecasting method. This is due to the rapid monthly changes in production using the seasonally adjusted data, which makes it very difficult for all the three methods to correctly predict sign of growth in these series.

\begin{minipage}[t]{.9\textwidth}                          
    \begin{center}
     \captionof{table}{Post-sample forecast accuracy measures for France}\label{fr2}
        \begin{adjustbox}{center, width=\columnwidth-80pt}  
\begin{tabular}{ccccccccccccccc}
\toprule
Series&Steps&\multicolumn{2}{r}{RMSE}& \multicolumn{2}{r}{RRMSE}& \multicolumn{2}{r}{DC}\\
\cmidrule(r){3-5} \cmidrule(r){6-7}\cmidrule(r){8-10}
&h&SSA&Boot SSA&GSSA&$\frac{Boot SSA}{SSA}$&$\frac{GSSA}{SSA}$&SSA&Boot SSA&GSSA\\
  \midrule
Electricity and gas	&h=1&	3.763	&	3.765	&	2.725	&	1.001	&	0.724	*&	0.575	&	0.575	&	0.582	\\
	&h=3&	4.450	&	4.457	&	2.850	&	1.002	&	0.640	*&	0.568	&	0.568	&	0.553	\\
	&h=6&	4.964	&	4.978	&	3.512	&	1.003	&	0.707	*&	0.519	&	0.519	&	0.574	\\
	&h=12&	5.612	&	5.641	&	4.694	&	1.005	&	0.836	*&	0.569	&	0.577	&	0.602	\\
\hline																	
Chemical	&h=1&	1.988	&	1.989	&	1.502	&	1.000	&	0.756	*&	0.512	&	0.512	&	0.465	\\
	&h=3&	3.009	&	3.010	&	2.411	&	1.000	&	0.801	*&	0.537	&	0.537	&	0.585	\\
	&h=6&	3.937	&	3.940	&	3.298	&	1.001	&	0.838	&	0.553	&	0.553	&	0.526	\\
	&h=12&	3.453	&	3.453	&	3.187	&	1.000	&	0.923	&	0.656	&	0.656	&	0.656	\\
\hline																	
Fabricate Metals	&h=1&	2.991	&	2.993	&	1.519	&	1.001	&	0.508	*&	0.636	&	0.636	&	0.636	\\
	&h=3&	5.497	&	5.502	&	1.886	&	1.001	&	0.443	*&	0.657	&	0.648	&	0.524	\\
	&h=6&	8.727	&	8.733	&	4.144	&	1.001	&	0.475	&	0.539	&	0.529	&	0.451	\\
	&h=12&	13.743	&	13.742	&	12.816	&	1.000	&	0.933	&	0.469	&	0.469	&	0.458	\\
\hline																	
Vehicle	&h=1&	6.504	&	6.505	&	3.891	&	1.000	&	0.598	*&	0.546	&	0.546	&	0.528	\\
	&h=3&	10.340	&	10.351	&	4.801	&	1.001	&	0.464	&	0.500	&	0.500	&	0.462	\\
	&h=6&	15.499	&	15.527	&	8.637	&	1.002	&	0.557	&	0.505	&	0.505	&	0.524	\\
	&h=12&	21.694	&	21.739	&	18.237	&	1.002	&	0.841	&	0.443	&	0.454	&	0.495	\\
\hline																	
Food Products	&h=1&	1.272	&	1.272	&	1.017	&	1.000	&	0.800	*&	0.659	&	0.659	&	0.659	\\
	&h=3&	1.352	&	1.349	&	1.024	&	0.998	&	0.757	*&	0.690	&	0.690	&	0.738	\\
	&h=6&	1.840	&	1.838	&	1.411	&	0.999	&	0.767	*&	0.615	&	0.615	&	0.667	\\
	&h=12&	2.542	&	2.525	&	2.341	&	0.993	&	0.921	*&	0.576	&	0.606	&	0.576	\\
\hline																	
Basic Metals	&h=1&	2.727	&	2.728	&	1.485	&	1.001	&	0.545	*&	0.621	&	0.621	&	0.646	\\
	&h=3&	4.977	&	4.982	&	1.841	&	1.001	&	0.370	&	0.553	&	0.553	&	0.541	\\
	&h=6&	7.775	&	7.780	&	3.750	&	1.001	&	0.482	&	0.513	&	0.519	&	0.500	\\
	&h=12&	11.707	&	11.705	&	11.357	&	1.000	&	0.970	&	0.507	&	0.507	&	0.560	\\
\hline																	
Electrical Machinery	&h=1&	2.978	&	2.980	&	1.496	&	1.001	&	0.502	&	0.610	&	0.610	&	0.610	\\
	&h=3&	5.136	&	5.144	&	1.986	&	1.002	&	0.487	&	0.485	&	0.485	&	0.476	\\
	&h=6&	8.333	&	8.345	&	4.829	&	1.002	&	0.579	&	0.510	&	0.510	&	0.550	\\
	&h=12&	13.749	&	13.761	&	12.616	&	1.001	&	0.917	&	0.479	&	0.489	&	0.500	\\
\hline																	
Machinery	&h=1&	3.893	&	3.896	&	1.445	&	1.001	&	0.371	&	0.438	&	0.438	&	0.438	\\
	&h=3&	7.020	&	7.030	&	1.829	&	1.002	&	0.260	&	0.408	&	0.408	&	0.466	\\
	&h=6&	11.576	&	11.593	&	5.422	&	1.001	&	0.468	&	0.420	&	0.420	&	0.520	\\
	&h=12&	19.474	&	19.494	&	17.892	&	1.001	&	0.918	&	0.436	&	0.436	&	0.447	\\
\hline																	
Summery	&h=1&	3.265	&	3.266	&	1.885	&	1.000	&	0.600	&	0.574	&	0.574	&	0.570	\\
	&h=3&	5.223	&	5.228	&	2.328	&	1.001	&	0.503	&	0.550	&	0.549	&	0.541	\\
	&h=6&	7.831	&	7.842	&	4.375	&	1.001	&	0.609	&	0.522	&	0.521	&	0.539	\\
	&h=12&	11.497	&	11.508	&	10.392	&	1.000	&	0.907	&	0.517	&	0.524	&	0.537	\\

\hline							            	
 \end{tabular}
 \end{adjustbox}
  \end{center}
    \center
\scriptsize{\textit{Note}:* indicates results are statistically significant at p=0.01 based on modified Diebold-Mariano test.
}
\end{minipage}
\\
\begin{minipage}[t]{.9\textwidth}                          
    \begin{center}
     \captionof{table}{Post-sample forecast accuracy measures for Germany} \label{ger2}
        \begin{adjustbox}{center, width=\columnwidth-80pt}  
\begin{tabular}{ccccccccccccccc}
\toprule
Series&Steps&\multicolumn{2}{r}{RMSE}& \multicolumn{2}{r}{RRMSE}& \multicolumn{2}{r}{DC}\\
\cmidrule(r){3-5} \cmidrule(r){6-7}\cmidrule(r){8-10}
&h&SSA&Boot SSA&GSSA&$\frac{Boot SSA}{SSA}$&$\frac{GSSA}{SSA}$&SSA&Boot SSA&GSSA\\
  \midrule
Electricity and gas	&h=1&	3.468	&	3.471	&	1.785	&	1.001	&	0.515	*&	0.500	&	0.500	&	0.537	\\
	&h=3&	4.642	&	4.643	&	3.635	&	1.000	&	0.783	*&	0.621	&	0.614	&	0.614	\\
	&h=6&	5.376	&	5.375	&	4.522	&	1.000	&	0.841	*&	0.612	&	0.612	&	0.620	\\
	&h=12&	5.724	&	5.704	&	5.244	&	0.997	&	0.916	*&	0.634	&	0.634	&	0.618	\\
\hline																	
Chemical	&h=1&	2.310	&	2.312	&	1.247	&	1.001	&	0.540	*&	0.607	&	0.607	&	0.509	\\
	&h=3&	3.608	&	3.614	&	2.691	&	1.002	&	0.746	*&	0.582	&	0.573	&	0.582	\\
	&h=6&	5.338	&	5.346	&	4.204	&	1.001	&	0.787	*&	0.551	&	0.533	&	0.551	\\
	&h=12&	7.498	&	7.521	&	6.700	&	1.003	&	0.894	&	0.545	&	0.554	&	0.535	\\
\hline																	
Fabricate Metals	&h=1&	2.489	&	2.493	&	1.541	&	1.002	&	0.619	*&	0.475	&	0.475	&	0.495	\\
	&h=3&	5.043	&	5.054	&	3.858	&	1.002	&	0.765	&	0.394	&	0.404	&	0.455	\\
	&h=6&	8.245	&	8.262	&	6.608	&	1.002	&	0.801	&	0.333	&	0.333	&	0.490	\\
	&h=12&	13.138	&	13.155	&	12.634	&	1.001	&	0.962	*&	0.378	&	0.378	&	0.444	\\
\hline																	
Vehicle	&h=1&	4.926	&	4.928	&	2.613	&	1.000	&	0.530	*&	0.710	&	0.710	&	0.661	\\
	&h=3&	7.200	&	7.211	&	4.593	&	1.002	&	0.638	*&	0.639	&	0.648	&	0.631	\\
	&h=6&	9.995	&	10.011	&	8.819	&	1.002	&	0.882	*&	0.588	&	0.597	&	0.529	\\
	&h=12&	14.205	&	14.234	&	13.265	&	1.002	&	0.934	*&	0.566	&	0.549	&	0.584	\\
\hline																	
Food Products	&h=1&	1.678	&	1.679	&	0.974	&	1.000	&	0.580	*&	0.708	&	0.708	&	0.547	\\
	&h=3&	1.833	&	1.835	&	1.551	&	1.001	&	0.846	&	0.635	&	0.635	&	0.635	\\
	&h=6&	1.952	&	1.958	&	1.406	&	1.003	&	0.720	&	0.693	&	0.703	&	0.624	\\
	&h=12&	2.464	&	2.481	&	2.081	&	1.007	&	0.845	&	0.611	&	0.611	&	0.589	\\
\hline																	
Basic Metals	&h=1&	3.778	&	3.781	&	1.691	&	1.001	&	0.448	*&	0.538	&	0.529	&	0.538	\\
	&h=3&	7.063	&	7.075	&	5.408	&	1.002	&	0.766	&	0.510	&	0.510	&	0.510	\\
	&h=6&	11.160	&	11.174	&	9.927	&	1.001	&	0.890	&	0.525	&	0.525	&	0.535	\\
	&h=12&	15.590	&	15.596	&	14.882	&	1.000	&	0.955	&	0.473	&	0.473	&	0.473	\\
\hline																	
Electrical Machinery	&h=1&	2.581	&	2.586	&	2.000	&	1.002	&	0.775	*&	0.436	&	0.436	&	0.505	\\
	&h=3&	4.947	&	4.962	&	3.158	&	1.003	&	0.638	&	0.364	&	0.364	&	0.505	\\
	&h=6&	8.315	&	8.339	&	6.218	&	1.003	&	0.748	&	0.354	&	0.354	&	0.448	\\
	&h=12&	13.749	&	13.781	&	11.944	&	1.002	&	0.869	&	0.411	&	0.411	&	0.444	\\
\hline																	
Machinery	&h=1&	3.714	&	3.716	&	2.952	&	1.001	&	0.795	*&	0.687	&	0.687	&	0.677	\\
	&h=3&	5.812	&	5.823	&	4.774	&	1.002	&	0.821	&	0.526	&	0.526	&	0.557	\\
	&h=6&	9.598	&	9.622	&	7.032	&	1.002	&	0.733	&	0.532	&	0.543	&	0.479	\\
	&h=12&	16.105	&	16.139	&	13.907	&	1.002	&	0.864	*&	0.466	&	0.466	&	0.523	\\
\hline																	
Summery	&h=1&	3.118	&	3.121	&	1.850	&	1.001	&	0.600	&	0.583	&	0.581	&	0.559	\\
	&h=3&	5.018	&	5.027	&	3.709	&	1.002	&	0.750	&	0.534	&	0.534	&	0.561	\\
	&h=6&	7.497	&	7.511	&	6.092	&	1.002	&	0.800	&	0.524	&	0.525	&	0.535	\\
	&h=12&	11.059	&	11.077	&	10.082	&	1.002	&	0.905	&	0.510	&	0.509	&	0.526	\\

\hline
\end{tabular}
        \end{adjustbox}
    \end{center}
  \center
\scriptsize{\textit{Note}:* indicates results are statistically significant at p=0.01 based on modified Diebold-Mariano test.
}
\end{minipage}
\\
\begin{minipage}[t]{.9\textwidth}                          
    \begin{center}
     \captionof{table}{Post-sample forecast accuracy measures for UK}\label{uk2}
        \begin{adjustbox}{center, width=\columnwidth-80pt}  
\begin{tabular}{ccccccccccccccc}
\toprule
Series&Steps&\multicolumn{2}{r}{RMSE}& \multicolumn{2}{r}{RRMSE}& \multicolumn{2}{r}{DC}\\
\cmidrule(r){3-5} \cmidrule(r){6-7}\cmidrule(r){8-10}
&h&SSA&Boot SSA&GSSA&$\frac{Boot SSA}{SSA}$&$\frac{GSSA}{SSA}$&SSA&Boot SSA&GSSA\\
\midrule
Electricity and gas	&h=1&	3.193	&	3.196	&	1.687	&	1.001	&	0.528	*&	0.453	&	0.453	&	0.469	\\
	&h=3&	4.479	&	4.484	&	2.783	&	1.001	&	0.621	*&	0.645	&	0.645	&	0.565	\\
	&h=6&	4.618	&	4.622	&	3.726	&	1.001	&	0.807	&	0.542	&	0.525	&	0.542	\\
	&h=12&	6.026	&	6.028	&	5.282	&	1.000	&	0.877	&	0.491	&	0.491	&	0.509	\\
\hline																	
Chemical	&h=1&	2.523	&	2.526	&	2.115	&	1.001	&	0.838	*&	0.563	&	0.563	&	0.594	\\
	&h=3&	3.860	&	3.872	&	3.024	&	1.003	&	0.783	*&	0.597	&	0.597	&	0.597	\\
	&h=6&	4.800	&	4.817	&	3.216	&	1.004	&	0.670	*&	0.508	&	0.508	&	0.542	\\
	&h=12&	5.786	&	5.807	&	4.822	&	1.004	&	0.833	*&	0.509	&	0.509	&	0.528	\\
\hline																	
Fabricate Metals	&h=1&	2.563	&	2.566	&	2.300	&	1.001	&	0.897	*&	0.516	&	0.516	&	0.594	\\
	&h=3&	3.966	&	3.976	&	3.312	&	1.002	&	0.835	&	0.548	&	0.548	&	0.516	\\
	&h=6&	5.106	&	5.118	&	3.936	&	1.002	&	0.771	*&	0.542	&	0.542	&	0.542	\\
	&h=12&	6.534	&	6.528	&	5.094	&	0.999	&	0.780	&	0.623	&	0.623	&	0.604	\\
\hline																	
Vehicle	&h=1&	6.161	&	6.167	&	2.412	&	1.001	&	0.392	*&	0.470	&	0.470	&	0.470	\\
	&h=3&	10.551	&	10.568	&	7.356	&	1.002	&	0.697	*&	0.484	&	0.484	&	0.500	\\
	&h=6&	13.939	&	13.972	&	11.147	&	1.002	&	0.800	*&	0.525	&	0.525	&	0.492	\\
	&h=12&	14.605	&	14.654	&	11.453	&	1.003	&	0.784	&	0.436	&	0.436	&	0.491	\\
\hline																	
Food Products	&h=1&	1.732	&	1.734	&	1.430	&	1.001	&	0.825	*&	0.575	&	0.575	&	0.575	\\
	&h=3&	2.339	&	2.341	&	1.498	&	1.001	&	0.640	*&	0.563	&	0.563	&	0.592	\\
	&h=6&	3.129	&	3.134	&	2.046	&	1.002	&	0.654	*&	0.544	&	0.544	&	0.574	\\
	&h=12&	4.866	&	4.868	&	4.083	&	1.001	&	0.839	&	0.452	&	0.452	&	0.452	\\
\hline																	
Basic Metals	&h=1&	5.698	&	5.703	&	4.140	&	1.001	&	0.727	*&	0.563	&	0.563	&	0.578	\\
	&h=3&	7.965	&	7.979	&	7.017	&	1.002	&	0.881	*&	0.565	&	0.565	&	0.500	\\
	&h=6&	10.114	&	10.134	&	6.979	&	1.002	&	0.690	*&	0.593	&	0.593	&	0.593	\\
	&h=12&	11.417	&	11.413	&	8.916	&	1.000	&	0.781	&	0.604	&	0.604	&	0.509	\\
\hline																	
Electrical Machinery	&h=1&	4.659	&	4.662	&	3.735	&	1.001	&	0.802	*&	0.516	&	0.516	&	0.516	\\
	&h=3&	6.392	&	6.402	&	4.585	&	1.002	&	0.717	*&	0.517	&	0.500	&	0.517	\\
	&h=6&	8.735	&	8.758	&	5.730	&	1.003	&	0.656	*&	0.421	&	0.456	&	0.474	\\
	&h=12&	11.113	&	11.154	&	9.122	&	1.004	&	0.821	*&	0.549	&	0.549	&	0.549	\\
\hline																	
Machinery	&h=1&	4.139	&	4.143	&	3.289	&	1.001	&	0.795	*&	0.443	&	0.443	&	0.514	\\
	&h=3&	6.160	&	6.169	&	3.378	&	1.001	&	0.548	*&	0.441	&	0.441	&	0.426	\\
	&h=6&	9.896	&	9.907	&	6.136	&	1.001	&	0.620	*&	0.492	&	0.492	&	0.508	\\
	&h=12&	15.146	&	15.158	&	12.161	&	1.001	&	0.803	*&	0.525	&	0.525	&	0.559	\\
\hline																	
Summery	&h=1&	3.834	&	3.837	&	2.639	&	1.001	&	0.726	&	0.512	&	0.512	&	0.539	\\
	&h=3&	5.714	&	5.724	&	4.119	&	1.002	&	0.716	&	0.545	&	0.543	&	0.527	\\
	&h=6&	7.542	&	7.558	&	5.364	&	1.002	&	0.708	&	0.521	&	0.523	&	0.533	\\
	&h=12&	9.436	&	9.451	&	7.617	&	1.001	&	0.815	&	0.524	&	0.524	&	0.525	\\

\hline							            	
\end{tabular}
        \end{adjustbox}
    \end{center}
          \center
\scriptsize{\textit{Note}:* indicates results are statistically significant at p=0.01 based on modified Diebold-Mariano test.
}
\end{minipage}

\section{Discussion and Conclusions}
This study has proposed a General Singular Spectrum Analysis (GSSA) model and compared its performance with the basic SSA and bootstrap SSA for forecasting seasonally adjusted monthly data on industrial production indicators in France, Germany and UK. We found strong evidence that the modified SSA technique with the general recurrent formula performs substantially better than SSA and Bootstrap SSA methods for these production series according to the conventional  RMSE criterion. Table \ref{Sumfo} presents the summary statistics of the RMSE, RRMSE, number of significant cases and percentage of the correct sign forecasts computed across all series and countries. The results indicate the superiority of using the general SSA technique for out-of-sample forecasting, with overall reduction of 28$\%$  according to RMSE criterion. The results also show that the improvement is 36$\%$ for one step-ahead forecast, $h$$=$1, decreasing to 13$\%$ as h increases to 12 months ahead.
Comparing GSSA forecasts with SSA, GSSA outperforms SSA significantly 22, 14, 12, and 9 times (out of 24 cases) at h$=$1,3,6 and 12 horizons respectively at 1$\%$ level. The last column in Table \ref{Sumfo} shows the number of significant cases and indicates that for all the horizons and across all the countries, GSSA outperforms SSA  significantly at 1$\%$ level in 60$\%$ of cases (57 out of 96 cases). The graph of the cumulative density function also confirms the findings, showing that the errors obtained by the general SSA are stochastically smaller than the errors obtained by other models for h$=$1,3 and 6.
Our study finds similar percentage of forecasts of the correct sign for all the horizons. This is the case within each country, as well as being true overall, with only slightly more than 50$\%$ of the correct sign for all the models. This, however, due to rapid monthly changes in the sign of growth for these seasonally adjusted series. Therefore, we found no clear advantage for forecasting the direction of  future monthly growth for these seasonally adjusted series by any of the three methods.

\begin{minipage}[t]{.9\textwidth}
    \begin{center}
     \captionof{table}{Summery statistics for out-of-sample forecasting accuracy measures.}\label{Sumfo}
        \begin{adjustbox}{center, width=\columnwidth-80pt}  
\begin{tabular}{ccccccccccccccc}
\toprule
Steps&\multicolumn{3}{l}{RMSE}& \multicolumn{3}{l}{RRMSE}& \multicolumn{2}{l}{DC}&\multicolumn{1}{l}{Sig. at}\\
\cmidrule(r){2-4} \cmidrule(r){5-6}\cmidrule(r){7-9}\cmidrule(r){10-10}
&SSA&Boot SSA&GSSA&$\frac{Boot SSA}{SSA}$&$\frac{GSSA}{SSA}$&SSA&Boot SSA&GSSA&$0.01$\\
\midrule
h=1&	3.406	&	3.408	&	2.125	&	1.001	&	0.642	&	0.556	&	0.556	&	0.556	&	22\\
h=3&	5.318	&	5.326	&	3.386	&	1.002		&	0.656	&	0.543	&	0.542	&	0.543	&	14		\\
h=6&	7.623	&	7.637	&	5.277	&	1.002		&	0.706	&	0.522	&	0.523	&	0.536	&	12		\\
h=12&	10.664	&	10.679	&	9.364	&	1.001		&	0.876	&	0.517	&	0.519	&	0.529	&	9		\\
\hline
Overall&	\bf 6.753	&\bf	6.762	&\bf	5.038	&\bf	1.001	&\bf	0.720	&\bf	0.535	&\bf	0.535	&\bf	0.541	&\bf	57		\\
\hline							            	
\end{tabular}
        \end{adjustbox}
    \end{center}
\end{minipage}
\newline
\newline
The results obtained in this study are promising indications that using a more general form for the recurrent formula in the forecast period would produce better forecasts. We are currently studying the forecasting performance of another recurrent formula given by :	
\begin{equation}
y_t=\phi_0(Y_{t-1})+\phi_1 y_{t-1}+\phi_2 y_{t-2}+\ldots+\phi_{L-1}y_{t-(L-1)}.
\label{SDMAR1}
\end{equation}

The difference between the original recurrent formula (\ref{AR}) used in basic SSA and this model is that (\ref{SDMAR1}) has an additional parameter, $\phi_0$ called “mean or adjustment parameter ”. In this model only $\phi_0$ is determined by the past values $\{y_{t-1},\ldots,y_{t-(L-1)}\}$ and updated recursively in the out-of-sample forecast period. The other parameters $\phi_1,\phi_2,\ldots,\phi_{L-1}$ remain constant in the forecast period, the same as those obtained by the optimal SSA approach within the sample period.
\newpage

\section*{References}

\bibliography{mybibfile}
\end{document}